\documentclass[preprint,12pt]{elsarticle}

\usepackage{amsmath}
\usepackage{amsfonts}
\usepackage{amsthm}
\usepackage{amssymb}
\usepackage{array}
\usepackage{blkarray}
\usepackage{framed}
\usepackage{booktabs}
\usepackage[normalem]{ulem}
\usepackage{float}
\usepackage{xcolor}

\def\R{{\mathbb R}}
\def\N{{\mathbb N}}
    
\def\Sp{{\mathbb S}}

\def\Shape{\mathcal{S}}
\def\D{\mathcal{D}}
\def\I{\mathcal{I}}

\def\argmin{{\operatorname{argmin}}}

\newcommand{\vol}{\operatorname{vol}}
\newtheorem*{remark}{Remark}

\usepackage{graphicx}
\usepackage{subcaption}
\usepackage{xcolor}
\usepackage{algorithm}
\usepackage{algpseudocode}
\usepackage{amsmath}
\usepackage{graphicx}
\usepackage{xcolor}
\usepackage{algorithm}
\usepackage{algpseudocode}

\newcommand{\red}[1]{\textcolor{black}{#1}}
\definecolor{revision}{RGB}{200, 0, 0} 
\newcommand{\rev}[1]{\textcolor{black}{#1}}

\journal{Computers in Biology and Medicine}

\begin{document}

\begin{frontmatter}



\title{A computational pipeline for clustering left atrial appendage morphology via elastic shape analysis}


\author[1,2]{Zan Ahmad} 
\author[2]{Minglang Yin}
\author[1]{Yashil Sukurdeep}
\author[1]{Noam Rotenberg}
\author[3]{Ritu Yadav}
\author[3]{Jenna Milstein}
\author[3]{Linh Thi My Tran}
\author[3]{Calvin O'Donnell}
\author[3]{Sarah Schumacher}
\author[3]{Craig Cronin}
\author[3]{Robert Weinstein}
\author[3]{Danish Iltaf Satti}
\author[3]{David Spragg}
\author[1]{Eugene Kholmovksi}
\author[1,2]{Natalia Trayanova}

\affiliation[1]{organization={Department of Applied Mathematics and Statistics, Johns Hopkins University},
            city={Baltimore}, 
            state={MD},
            country={USA}}
\affiliation[2]{organization={Department of Biomedical Engineering, Johns Hopkins University},
            city={Baltimore}, 
            state={MD},
            country={USA}}
\affiliation[3]{organization={Johns Hopkins Hospital},
            city={Baltimore}, 
            state={MD},
            country={USA}}

\begin{abstract}
Morphological variations in the left atrial appendage (LAA) are associated with different levels of ischemic stroke risk for patients with atrial fibrillation (AF). Studying LAA morphology can elucidate mechanisms behind this association and lead to the development of advanced stroke risk stratification tools. However, current categorical descriptions of LAA morphologies are qualitative in nature, and inconsistent across studies, which impedes advancements in our understanding of stroke pathogenesis in AF. To mitigate these issues, we introduce a quantitative pipeline that combines elastic shape analysis with unsupervised learning for the categorization of LAA morphology in AF patients. We demonstrate that our method reliably clusters LAAs based on their geometric features, and thus provides an avenue to overcome the limitations of current qualitative LAA categorization systems. 
\end{abstract}


\end{frontmatter}


\section{Introduction}
\label{intro}
Atrial fibrillation (AF), the most common cardiac arrhythmia, is characterized by irregular contractions of the atrial chambers~\cite{january20192019}. It represents an emerging global health crisis, affecting $2-3\%$ of individuals worldwide, with its prevalence expected to increase 2.5-fold in the next 40 years~\cite{pistoia2016epidemiology}. This rhythm disorder is associated with blood stagnation and clot formation in the left atrial appendage (LAA) -- a tubular pouch attached to the main body of the left atrium~\cite{yang2023observational}. A devastating complication of AF occurs when clots detach from the LAA and travel to the brain, causing vessel blockage and leading to ischemic stroke~\cite{wolf1991atrial}. The risk of stroke in AF patients is five times that of a healthy individual~\cite{piccini2016preventing}.

Numerous studies have shown that LAA morphology is correlated with different levels of ischemic stroke risk \cite{tian2020morphological,bieging2021statistical,burrell2013usefulness,di2012does}. Despite this evidence, stroke risk stratification metrics for AF patients typically overlook geometric considerations of the LAA \cite{harb2018p5142}. Furthermore, these metrics often inaccurately determine which patients should receive oral anticoagulants, which can cause severe side effects such as internal bleeding and intracranial hemorrhage \cite{harb2018p5142,steiner2006intracerebral,chen2019cha2ds2,zhang2021interpretation,dudzinska2022association,sun2023finding,lodzinski2020trends}. Additionally, in current LAA shape classification systems, the ``chicken-wing" appendage is known to exhibit lower stroke prevalence compared to other categories (e.g., windsock, cauliflower, cactus)~\cite{di2012does,fang2022stroke}. However, the high inter-observer variability in these qualitative classification systems has led to inconsistent definitions of LAA morphology categories across different studies and patient populations \cite{smit2021anatomical}. Consequently, mixed conclusions have been documented regarding the association between LAA morphologies and ischemic stroke risk, hindering the incorporation of LAA shape information into stroke risk assessments \cite{yaghi2020left}. Therefore, an objective and quantitative framework for rigorous LAA shape categorization is imperative.

\subsection{Related work}
\label{ssec:related_work}
Several studies have attemped to quantify LAA shape using simple metrics and functional data~\cite{bieging2021statistical,zingaro2024comprehensive,zingaro2023po,slodowska2021morphology}, but only a few prior works have leveraged detailed shape analysis frameworks. 

For instance, Slipsager et al. developed an approach for the statistical shape analysis and clustering of LAA meshes based on non-rigid volumetric registration of signed distance fields, and Juhl et al. further extended this work~\cite{slipsager2019statistical,juhl2024signed}. These quantitative pipelines were successful at identifying two distinct LAA shape clusters corresponding to the so-called chicken-wing and non-chicken-wing morphologies. 

Furthermore, in a recent investigation, Lee et al. used a landmark-based statistical shape analysis approach for LAA shape categorization~\cite{lee2024preserving}. In particular, they used dedicated workflows from the \texttt{ShapeWorks} software~\cite{cates2017shapeworks} to cluster LAA meshes across a range of mesh resolutions. \rev{Their approach involved globally aligning LAA meshes using the Super4PCS algorithm, followed by iterative local refinement with the Iterative Closest Point (ICP) algorithm and multiview registration. A particle-based statistical shape modeling approach was then implemented in \texttt{ShapeWorks}, where particle correspondences were iteratively optimized across the dataset to capture shape variability. The final statistical shape model was analyzed using principal component analysis (PCA) to quantify morphological differences. Their multi-scale analysis demonstrated that higher-resolution meshes improved shape categorization by preserving finer anatomical details of complex and diverse LAA morphologies.}

The works of~\cite{slipsager2019statistical,juhl2024signed} face an important limitation, as they require point-to-point correspondences between LAA meshes (i.e., parametrized 3D surfaces) for the ensuing statistical shape analysis. Establishing these point correspondences can be challenging, as it involves solving a registration problem which often incurs high computational costs, and can yield unsatisfactory results with complex geometric data (such as LAA meshes containing fine-grained anatomical details). Indeed, in \cite{juhl2024signed}, the authors mention that they chose to refine point correspondences on the decoupled LAAs due to large differences in size and complexity between the LAA and the remaining left atrium, making it difficult to maintain good correspondence over the entire mesh. Even state-of-the-art methods for finding point correspondences with 3D geometric data, such as the functional maps framework, struggle when used with complex geometric data as they require a good prior selection of landmarks~\cite{ovsjanikov2012functional}.

\subsection{Contributions}
\label{ssec:contributions}
In this paper, we propose a novel computational pipeline for LAA morphology categorization which combines automated atrial segmentation, mesh generation, \emph{elastic shape distance} computations, and unsupervised clustering. Using a set of 3D LAA meshes from 20 AF patients, we compute and leverage pairwise elastic distances to: (a) quantify the similarity between pairs of shapes by calculating the Riemannian energy required to deform one of them into the other, and (b) cluster the meshes into distinct categories based on these shape similarities. Our pipeline has the ability to operate on unparametrized surfaces, which circumvents the requirement of point-to-point correspondences between LAA meshes, and thus renders the pipeline readily implementable in clinical workflows.

\section{LAA categorization pipeline}
\label{sec:methods}
In what follows, we describe the data used in this article, provide theoretical background on elastic shape analysis (ESA), and detail our computational approach for clustering the LAA meshes.

\subsection{Data description and preparation}
\label{ssec:data_description}

\begin{figure}[b]
    \centering
    \includegraphics[width=\textwidth]{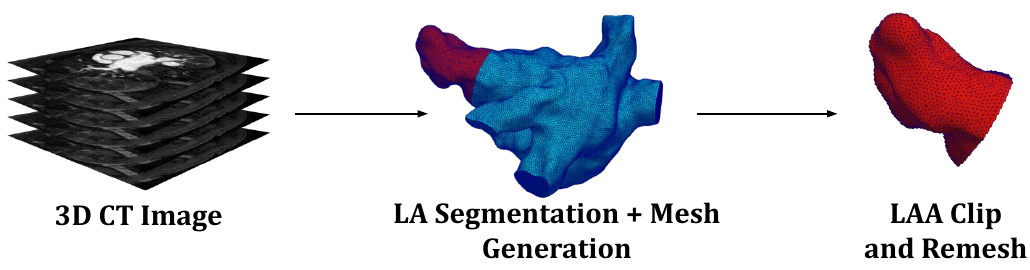}
    \caption{Schematic summary of the pre-processing pipeline: From 3D CT images, we automatically segment the LA using a deep learning approach, and generate a triangulated 3D mesh. Then the LAA (red) is clipped from the LA body (blue) and re-meshed.}
    \label{fig:LA-seg}
\end{figure}

A dataset of 20 de-identified, contrast-enhanced computed tomography (CT) scans of AF patients was used in this study. These images were collected from the Johns Hopkins Hospital (JHH) AF ablation registry from 2014 to 2022. The patients have drug-refractory AF and were documented to have undergone catheter ablation. CT images were acquired using a commercially available 320-detector CT scanner (Acquilion ONE, Canon Medical Systems) at JHH. This study complies with all relevant ethical regulations and was approved by the Johns Hopkins Medicine Institutional Review Boards (JHM IRBs). \rev{We excluded patients with left atrial appendage occlusion and poor CT quality in this study.}

We adapted our in-house deep learning model to automate left atrium (LA) segmentation on the CT scans~\cite{lefebvre2022lassnet}. The model has a cascade structure, where a region of interest is first cropped and followed by a convolutional neural network (CNN) that performs segmentation on the cropped area. All CNN-segmented scans were examined manually by a medical imaging expert for quality check. From the LA segmentations, we isolated the LAA from the LA chamber using \texttt{Paraview}'s clip tool, such that the ostium (selected to be aligned with the Coumadin ridge) was orthogonal to the initial protrusion of the appendage from the main body of the chamber, based on the LAA segmentation approach in \cite{schluchter2019vascular}. We then re-meshed the clipped LAA to yield the fully pre-processed geometry for shape analysis, see Figure~\ref{fig:LA-seg}.

\subsection{Elastic shape analysis}
\label{ssec:elastic_shape_analysis}
The ``shapes" of interest to us in this article are the LAA meshes, which we model mathematically (from a continuous viewpoint) as surfaces immersed in $\R^3$. As such, our approach for LAA shape comparison and categorization fundamentally relies on a quantitative measure of similarity (i.e., a distance) between surfaces. The framework of Riemannian shape analysis is particularly well-suited for defining such notions. At a high level, distances between pairs of geometric objects (such as LAA meshes) correspond to the minimal amount of `energy' required to morph one of the objects into the other via a combination of non-rigid geometric transformations, such as bending, stretching and shearing~\cite{hartman2023elastic,bauer2012sobolev,jermyn2017elastic,beg2005computing}. In what follows, we outline the theoretical underpinnings of this framework.

Let us start by defining a \textit{parametrized immersed surface} in $\R^3$, which refers to a smooth mapping $q \in C^{\infty}(M, \R^3)$, whose differential $dq$ is injective at every point of the parameter space $M$, which is a $2$-dimensional compact manifold (possibly with boundary) whose local coordinates are denoted by $(u,v) \in \R^2$. For example, $M$ can be a compact domain of $\R^2$ if one is working with open LAA meshes, or the sphere $\Sp^2$ in the setting of closed LAA meshes. The set of all parametrized surfaces, denoted by $\I$, is itself an infinite-dimensional manifold, where the tangent space at any $q \in \I$, denoted $T_q \I$, is given by $C^{\infty}(M,\R^3)$. Any tangent vector $h\in T_q\I$ can be thought of as a vector field along the surface $q$. 

\begin{figure}[htbp]
    \centering
    \includegraphics[width=.9\textwidth]{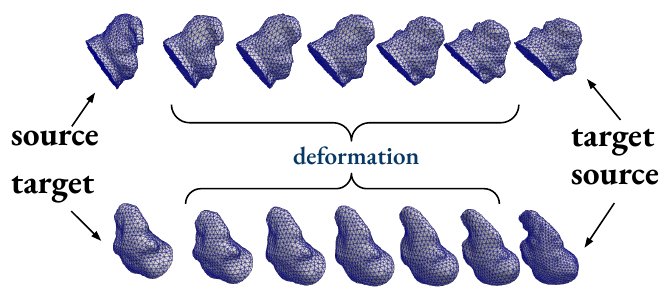}
    \caption{\rev{\textbf{Top:} An optimal deformation between a source ($q_0$) and target surface ($q_1$), computed by solving the relaxed-matching problem in Equation (\ref{eq:relaxed_matching}). This illustrates the optimal path in shape space minimizing the Riemannian energy which can be used in Equation (\ref{eq:elastic_distance_approx}) to approximate the distance in Equation (\ref{eq:shape_space_distance}). \textbf{Bottom:} The same geodesic shown from an alternative, rotated view, visualizing the smooth deformation governed by the Riemannian metric $G$ (Equation~\eqref{eq:metric_inv}).}}
    \label{fig:deformation}
\end{figure}

The key ingredient in Riemannian shape analysis is to equip the manifold $\I$ with a Riemannian metric $G$, which in our case refers to a family of inner products $G_q: T_q \I \times T_q \I \to \R$ that varies smoothly with respect to $q \in \I$. Indeed, any Riemannian metric $G$ on $\I$ induces a (pseudo) distance on this space, which is given for any two parametrized surfaces $q_0, q_1 \in \I$ by
\begin{equation}
\label{eq:geodesic_dist_parametrized}
d_G(q_0,q_1)^2=\underset{q(\cdot)} {\operatorname{inf}} \int_0^1 G_{q(t)}(\partial_t q(t),\partial_t q(t))~dt,
\end{equation}
where the infimum is taken over the space of all paths of immersed surfaces connecting $q_0$ and $q_1$, i.e., over all $q(\cdot) \in C^{\infty}([0, 1], \I)$ such that $q(0)=q_0$ and $q(1)=q_1$. Note that $\partial_t q(t)$ denotes the derivative of this path with respect to time $t$. We refer to minimization problem~\eqref{eq:geodesic_dist_parametrized} as the \emph{parametrized matching problem}, where $q_0$ and $q_1$ are called the \emph{source} and \emph{target} surfaces respectively, with $d
_G$ being known as the \emph{geodesic distance} between $q_0$ and $q_1$, and where minimizing paths (if they exist) are termed \emph{geodesics}. The functional being minimized in~\eqref{eq:geodesic_dist_parametrized} is the \emph{Riemannian energy} of the path $q(\cdot)$, and thus, geodesics can be interpreted as optimal deformations of minimal `energy' between $q_0$ and $q_1$ -- see Figure~\ref{fig:deformation} for an illustration. In a Riemannian setting, geodesics and distances obtained via~\eqref{eq:geodesic_dist_parametrized} form the basis of frameworks for the comparison and statistical shape analysis of surfaces, see~\cite{hartman2023elastic,su2020shape,bauer2021numerical}.

Nevertheless, we need to look beyond the setting of parametrized surfaces, i.e., geometric data with known point-to-point correspondences. Indeed, our goal is to develop a method for the comparison of LAA morphologies on datasets of unregistered LAA meshes, independently of how they are parametrized (or sampled/discretized). To this end, we introduce the \textit{reparametrization group} $\D$, i.e., the group of all diffeomorphisms of the parameter space $M$ (i.e., smooth and bijective maps $\varphi \in C^{\infty}(M)$ with a smooth inverse). For any surface $q \in \I$ and $\varphi \in \D$, we say that $q \circ \varphi \in \I$ is a \textit{reparametrization} of $q$ by $\varphi$. The discrete analogue of a reparametrization of a surface is a re-meshing of a discretized (e.g. triangulated) 3D mesh.

Since we seek geodesic distances between shapes regardless of how they are parametrized, let us introduce the \emph{shape space of unparametrized surfaces}, defined as the quotient space of parametrized surfaces modulo the reparametrization group $\Shape = \I / \D$. This space consists of equivalence classes $[q] = \{ q \circ \varphi | \varphi \in \D \}$ comprised of all reparametrizations of a given surface $q \in \I$. To obtain a distance on the shape space, we require a Riemannian metric $G$ that is invariant under the action of the aforementioned reparametrization group $\mathcal D$, i.e., we require
\begin{equation}
\label{eq:metric_inv}
G_q(h,k)=G_{q\circ\varphi}(h\circ\varphi,k\circ\varphi)    
\end{equation}
for all $q\in \I$, $h,k\in T_q\I$ and $\varphi \in \mathcal D$. The field of \emph{elastic shape analysis} (ESA) is a particularly powerful source of Riemannian metrics satisfying the reparametrization-invariance property above~\cite{bauer2010sobolev,jermyn2017elastic}. 

In this paper, we shall focus on \emph{second-order Riemannian Sobolev metrics ($H^2$-metrics)}, which have been shown to possess several desirable properties, including reparametrization-invariance, and which have successfully been used in applications with unregistered 3D geometric data~\cite{hartman2023elastic,bauer2010sobolev,hartman2023bare,hartman2023basis}. For all $q\in \I$ and $h,k\in T_q\I$, the family of $H^2$-metrics is defined as follows:
\begin{equation}
\label{eq:H2metric}
\begin{aligned}
&G_q(h,k)=\int_M\bigg( a_0 \langle h,k \rangle +
a_1 g_q^{-1}(dh_m,dk_m)\\&\qquad\qquad +b_1g_q^{-1}(dh_+,dk_+)+ c_1g_q^{-1}(dh_\bot,dk_\bot)\\&\qquad\qquad+ d_1 g_q^{-1}(dh_0,dk_0)
+a_2 \langle\Delta_q h,\Delta_q k\rangle\bigg)\vol_q,
\end{aligned}
\end{equation}
where $a_0, a_1, b_1, c_1, d_1$ and $a_2$ are non-negative weighting coefficients for the different zeroth, first and second order terms in the metric. In the above, $dh$ is the vector valued one-form on $M$ given by the differential of $h$, i.e., a map from $TM$ to $\R^3$ which can be viewed as a $3 \times 2$ matrix field on $M$ (in a given coordinate system). Meanwhile, $g_q = q^*\langle \cdot , \cdot \rangle$ is the pullback of the Euclidean metric on $\mathbb R^3$, which can be represented as a $2 \times 2$ symmetric positive definite matrix field on $M$, implying that $g_q^{-1}(dh,dh)=\operatorname{tr}(dh~g_q^{-1}~dh^T)$. Finally, the second-order term involves the Laplacian $\Delta_q$ induced by $q$, which can be written in coordinate-form as $\Delta_qh= \operatorname{det}(g_q)^{-\frac{1}{2}}\partial_u (\operatorname{det}(g_q)^{\frac{1}{2}}~g_q^{uv}\partial_v h)$, and $\vol_q = |q_u \times q_v|dudv$ is the volume measure induced by $q$. 

The first-order term in the metric is split as a result of the orthogonal decomposition of $dh$ into the sum of $dh_m, dh_+, dh_\perp$ and $dh_0$, whose precise definitions are given in~\cite{hartman2023elastic}. When splitting the metric in this way, the terms weighted by $a_1, b_1$ and $c_1$ in~\eqref{eq:H2metric} correspond to shearing, stretching and bending energies induced by the deformation field $h$ respectively. Therefore, via the selection of the weighting coefficients, the class of invariant $H^2$-metrics provides us with the flexibility to emphasize or penalize different types of geometric deformations when computing geodesics. This is particularly useful when working with LAA meshes, as it allows us to incorporate prior knowledge about LAA morphologies when selecting parameters to compute distances, in turn helping us obtain meaningful distances for downstream tasks such as clustering.

Given a reparametrization-invariant Riemannian metric $G$, such as an $H^2$-metric~\eqref{eq:H2metric}, its associated geodesic distance function $d_G$ from~\eqref{eq:geodesic_dist_parametrized} descends to a distance $d_{\Shape}$ on the quotient shape space $\Shape$. For any pair of unparametrized surfaces $[q_0],[q_1] \in \Shape$, this distance is defined by the following \emph{unparametrized matching problem}:
\begin{equation}
    \label{eq:shape_space_distance}
    d_{\Shape}([q_0],[q_1])^2 = \underset{\varphi \in \D}{\inf}~d_G(q_0, q_1 \circ \varphi)^2 =  \underset{\left( \varphi,~q(\cdot) \right)}{\inf} \int_0^1 G_{q(t)}(\partial_t q(t),\partial_t q(t))dt.
\end{equation}
Here, the minimization occurs jointly over paths of surfaces $q(\cdot)$ from $q_0$ (with a fixed parametrization) to $[q_1]$, and over all reparametrizations $\varphi \in \D$ of $q_1$, i.e., over paths such that $q(0)=q_0$ and $q(1) = q_1 \circ \varphi$.

In practice, solving~\eqref{eq:shape_space_distance} numerically is challenging. Indeed, it is relatively straightforward to discretize the $H^2$-metric and Riemannian energy functional by considering surfaces discretized as triangulated meshes~\cite{hartman2023elastic}, which allows us to frame the minimization over paths of surfaces as a standard finite-dimensional optimization problem. Yet, optimizing over reparametrizations of $q_1$ is challenging as the discretization of the diffeomorphism group $\D$ and its action on surfaces is not straightforward, see~\cite{jermyn2017elastic,su2020shape}. This motivates the need for alternative numerical techniques for solving~\eqref{eq:shape_space_distance}. 

One such approach, recently proposed in~\cite{hartman2023elastic,bauer2021numerical}, deals indirectly with the minimization over reparametrizations of $q_1$ by
instead relaxing the end time constraint (i.e., only requiring $q(1) \approx q_1 \circ \varphi$) using a \emph{data attachment term} $\Gamma([q(1)], [q_1])$ that is independent of the parametrizations of either $q(1)$ or $q_1$. Broadly speaking, this approach involves solving the \emph{relaxed matching problem} below:
\begin{equation}
    \label{eq:relaxed_matching}
    \inf \left\{ \int_0^1 G_{q(t)}(\partial_t q(t),\partial_t q(t)) dt + \lambda \Gamma([q(1)], [q_1]) \right\},
\end{equation}
where the minimization occurs over paths of surfaces \mbox{$q(\cdot) \in C^{\infty}([0,1], \I)$} that satisfy the initial constraint $q(0) = q_0$ \emph{only}, and where $\Gamma([q(1)], [q_1])$ is a discrepancy term between the endpoint of the path $q(1)$ and the true target surface $q_1$, with \mbox{$\lambda > 0$} being a balancing parameter. In particular, this approach allows us to entirely bypass the hurdle of optimizing over the diffeomorphism group $\D$. In this paper, we follow the approach of past works with unregistered surface data~\cite{bauer2021numerical,hartman2023elastic,hartman2023bare} and define $\Gamma$ specifically as a kernel metric on the space of varifolds. For concision, we omit the technical details behind the construction of these so-called \emph{varifold fidelity metrics} in this paper, and instead refer interested readers to the works of~\cite{kaltenmark2017general,charon2013varifold} for further details.

To bring everything together, our approach for computing pairwise distances to compare and categorize LAA morphologies involves discretizing the LAA geometries as triangulated 3D meshes, and numerically solving relaxed matching problem~\eqref{eq:relaxed_matching} -- in which $G$ is an $H^2$-metric~\eqref{eq:H2metric} and $\Gamma$ is a varifold fidelity metric, as defined in~\cite{hartman2023elastic} -- via the L-BFGS algorithm. 

For clarity, in the results for LAA shape categorization presented in Section~\ref{sec:results}, the distances reported in Figure~\ref{fig:esaDmatrix} are estimates $\Tilde{d}_{\Shape}([q_0], [q_1])$ for the true elastic shape distance between a given a pair of unparametrized LAA meshes \mbox{$[q_0], [q_1] \in \Shape$}. These distance estimates are given by

\begin{equation}
    \label{eq:elastic_distance_approx}
    \Tilde{d}_{\Shape}([q_0], [q_1])^2 \doteq \int_0^1 G_{q(t)}(\partial_t q(t),\partial_t q(t))dt,
\end{equation}
where the integral above (i.e., the Riemannian energy) is evaluated at an optimal path of surfaces $q(\cdot)$ obtained by solving~\eqref{eq:relaxed_matching}.

\subsection{Clustering the LAA meshes}
\label{ssec:clustering_methods}
Equipped with pairwise elastic distances for our set of LAA meshes, our approach  for LAA shape categorization involves the following steps:

\begin{itemize}
    \item \textbf{Projection into Euclidean space:} First, we project the unparame-trized LAA meshes from the shape space \(\Shape\) into Euclidean space \(\mathbb{R}^m\) using multidimensional scaling (MDS), see~\ref{app:mds} for details.
    \item \textbf{Euclidean distance computation:} We then compute Euclidean distances between all pairs of projected LAA meshes in \(\mathbb{R}^m\).
    \item \textbf{Clustering in Euclidean space:} Finally, we apply standard algorithms to cluster the projected data points in $\R^m$. We choose \(K\)-means and agglomerative (hierarchical) clustering with average linkage as illustrative algorithms, see~\ref{app:kmeans} and~\ref{app:agglo_clustering} respectively for details.
\end{itemize}

Our clustering approach is motivated by the anatomical similarities and constrained variations in LAA morphology across subjects~\cite{tenenbaum2000global}. Indeed, due to their constrained geometric variability, the set of all LAA shapes resides on a relatively `small' neighborhood of the shape space of surfaces $\Shape$. As such, the subspace of all LAA shapes is likely to have an inherently low-dimensional structure. Consequently, methods such as MDS allow us to project the LAA meshes as points in low-dimensional Euclidean space while faithfully preserving information about their pairwise distances in the original shape space. Indeed, as evidenced by our empirical results presented in the next section, projecting the LAA meshes into two-dimensional Euclidean space $\R^2$ using MDS yields a projection where the Euclidean distance between the projected points seems to capture the shape dissimilarities between the LAA meshes rather fittingly (see Figure~\ref{fig:distances}). In turn, this allows us to cluster LAA meshes using their corresponding projections by employing standard clustering algorithms that operate in Euclidean space, which comes with the added benefit of lower computational costs, as the clustering algorithms operate on low-dimensional (projected) data points.

\subsection{Framework Summary}
\label{alg:laa_pipeline}
\begin{algorithm}[H]
\caption{\red{LAA Categorization Pipeline}}
\label{alg:laa_pipeline}
\begin{algorithmic}[1]
\Require \red{3D CT scans \( I = \{I_1, I_2, \ldots, I_n\} \)}
\Ensure \red{Cluster assignments \( C = \{c_1, c_2, \ldots, c_n\} \)}

\State \red{\textbf{Input:} CT scans \( I \)}

\State \red{\textbf{1. Data Preparation}}
\State \red{Segment left atrium (LA) using a CNN: \( S = f_{\text{CNN}}(I) \)}
\State \red{Isolate LAA: \( L = g(S) \) via clipping and re-meshing}

\State \red{\textbf{2. Elastic Shape Analysis (ESA)}}
\State \red{Model each \( L_i \) as \( q_i: M \to \mathbb{R}^3 \)}
\State \red{Compute approximation of the pairwise elastic distances:}
\State \red{\[
 d_{\Shape}([q_i], [q_j])^2 = \inf_{q, \varphi} \int_0^1 G_{q(t)}(\partial_t q, \partial_t q) \, dt
\]}
\State \red{using L-BFGS with varifold fidelity \( \Gamma \).}

\State \red{\textbf{3. Dimensionality Reduction}}
\State \red{Apply MDS: \( X = \text{MDS}(d_{\Shape}) \) with \( X \in \mathbb{R}^{n \times m} \)}

\State \red{\textbf{4. Clustering}}
\State \red{Cluster \( X \) using either:}
\begin{itemize}
    \item \red{$K$-means: \( C = \text{KMeans}(X, K) \)}
    \item \red{Agglomerative clustering: \( C = \text{Agglo}(X) \)}
\end{itemize}

\State \red{\textbf{Output:} Cluster assignments \( C \)}

\end{algorithmic}
\end{algorithm}









\section{LAA clustering results and analysis}
\label{sec:results}
We now present the results obtained from applying our clustering pipeline on our dataset of 20 LAA meshes described in Section~\ref{ssec:data_description}.

\subsection{Distance computations}
\label{ssec:dist_computations}
\begin{figure}[htbp]
  \centering
  \begin{subfigure}[b]{0.495\textwidth}
      \centering
      \includegraphics[width=\textwidth]{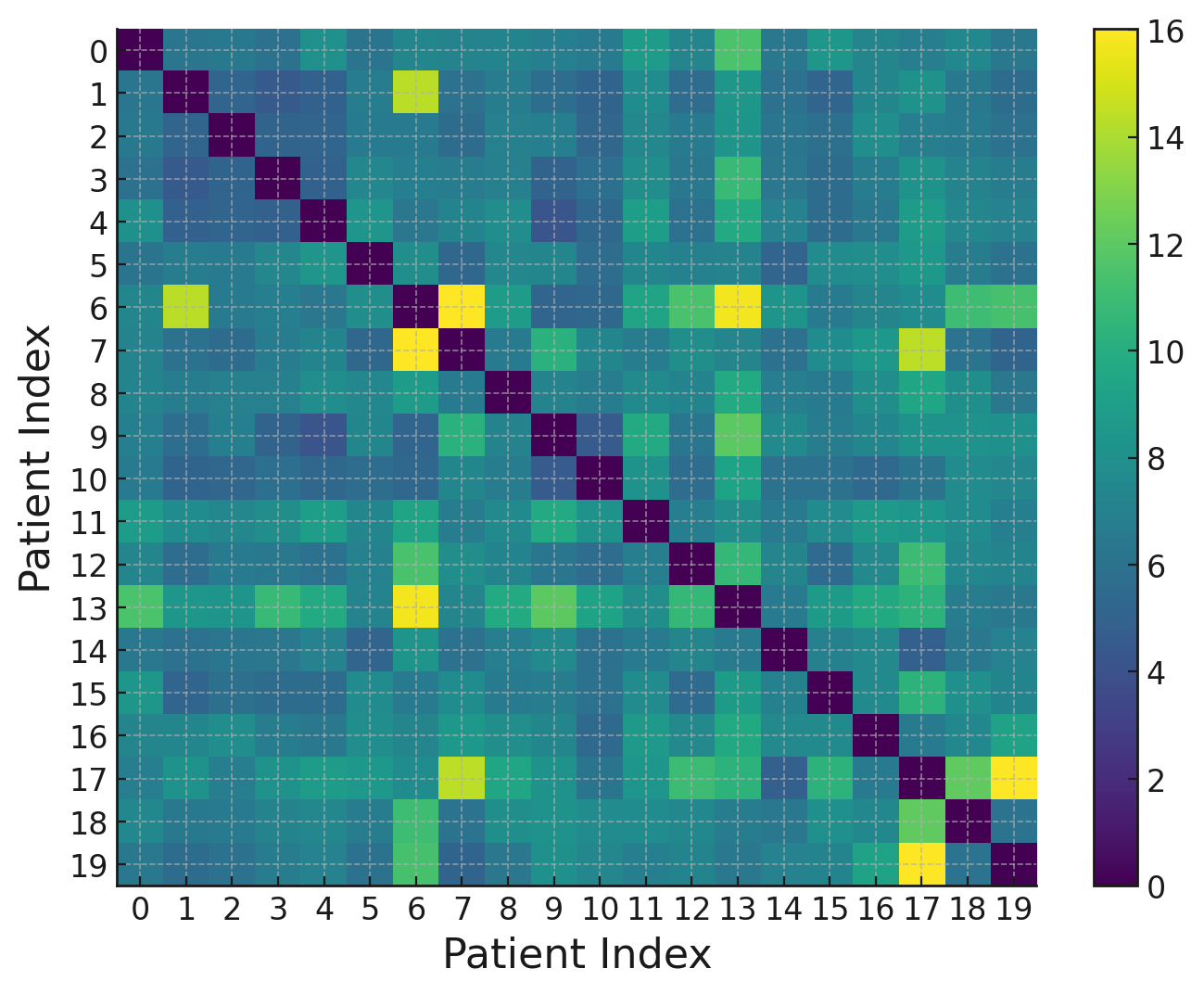}
      \caption{Elastic distances.}
      \label{fig:esaDmatrix}
  \end{subfigure}
  \begin{subfigure}[b]{0.495\textwidth}
      \centering
      \includegraphics[width=\textwidth]{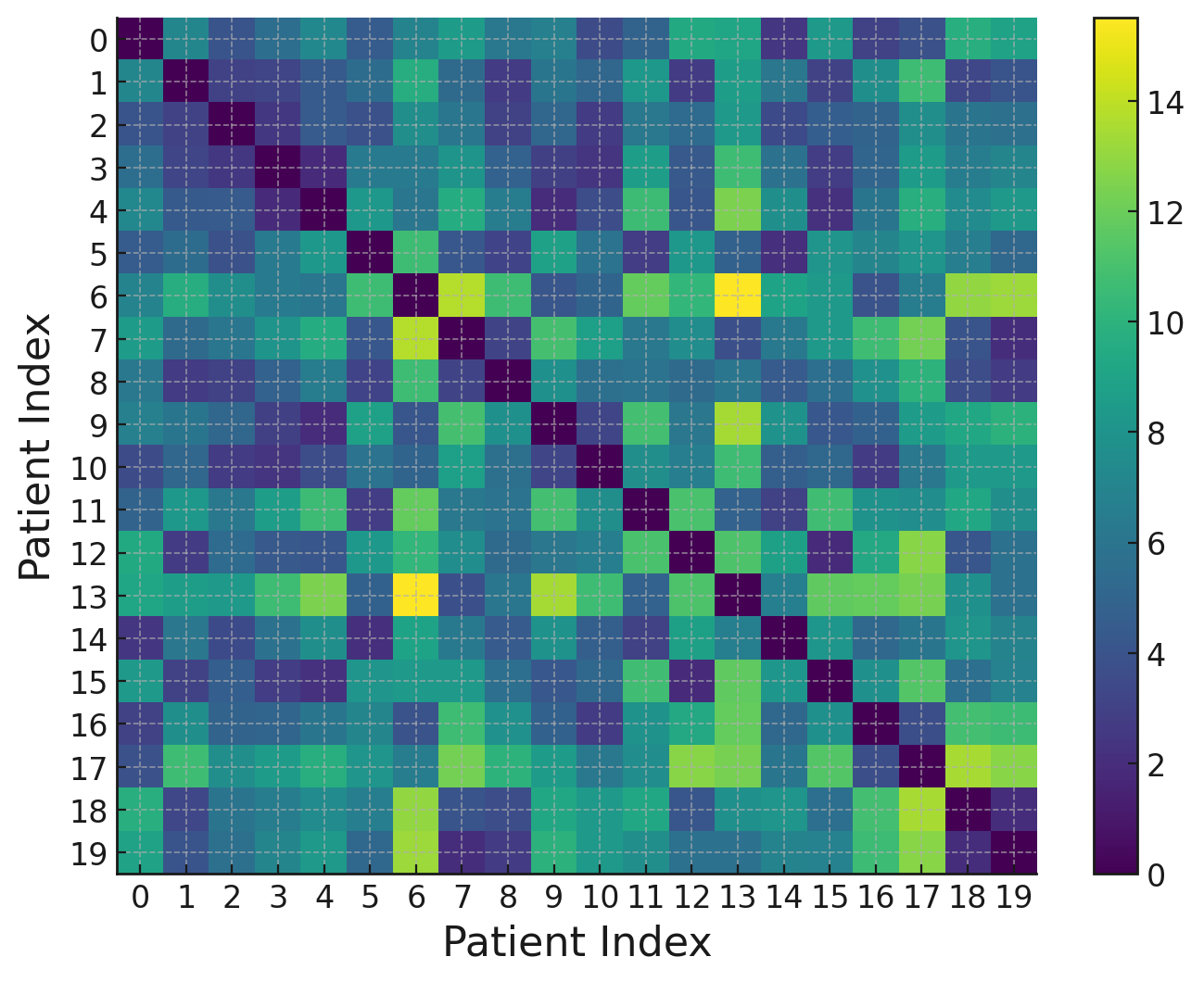}
      \caption{Euclidean distances after MDS projection.}
      \label{fig:mdsDmatrix}
  \end{subfigure}
  \caption{\rev{\textbf{Left (a):} Symmetric matrix representing pairwise \emph{elastic distances} for our cohort of 20 LAA meshes. These distances are calculated using Equation (\ref{eq:elastic_distance_approx}), by plugging in a geodesic path $q(\cdot)$ obtained by solving the relaxed matching problem described in Equation \ref{eq:relaxed_matching}. \textbf{Right (b):} Symmetric matrix representing pairwise \emph{Euclidean distances} between the projections of the LAA meshes in $\mathbb{R}^2$ obtained via MDS (see~\ref{app:mds}). The structural similarity between the matrices indicates that MDS effectively preserves the elastic distance relationships in a lower-dimensional Euclidean space.}}
  \label{fig:distances}
\end{figure}

We begin with Figure~\ref{fig:distances}, where we report the distances computed between all pairs of LAA meshes from the cohort. More specifically, Figure~\ref{fig:esaDmatrix} displays the pairwise elastic distances for our LAA cohort, which were computed by solving the relaxed matching problem~\eqref{eq:relaxed_matching}. Meanwhile, Figure~\ref{fig:mdsDmatrix} depicts the pairwise Euclidean distances between the MDS projections of the LAA meshes into $\R^2$. The key observation is that the structure of the two distance matrices is similar, which implies that MDS yields a projection of the LAA meshes in $\R^2$ that faithfully preserves information about their relative pairwise distances in the original shape space.

\subsection{LAA clusters}
\label{ssec:laa_clusters}
\begin{figure}[htbp]
        \centering
        \includegraphics[width=.76\textwidth]{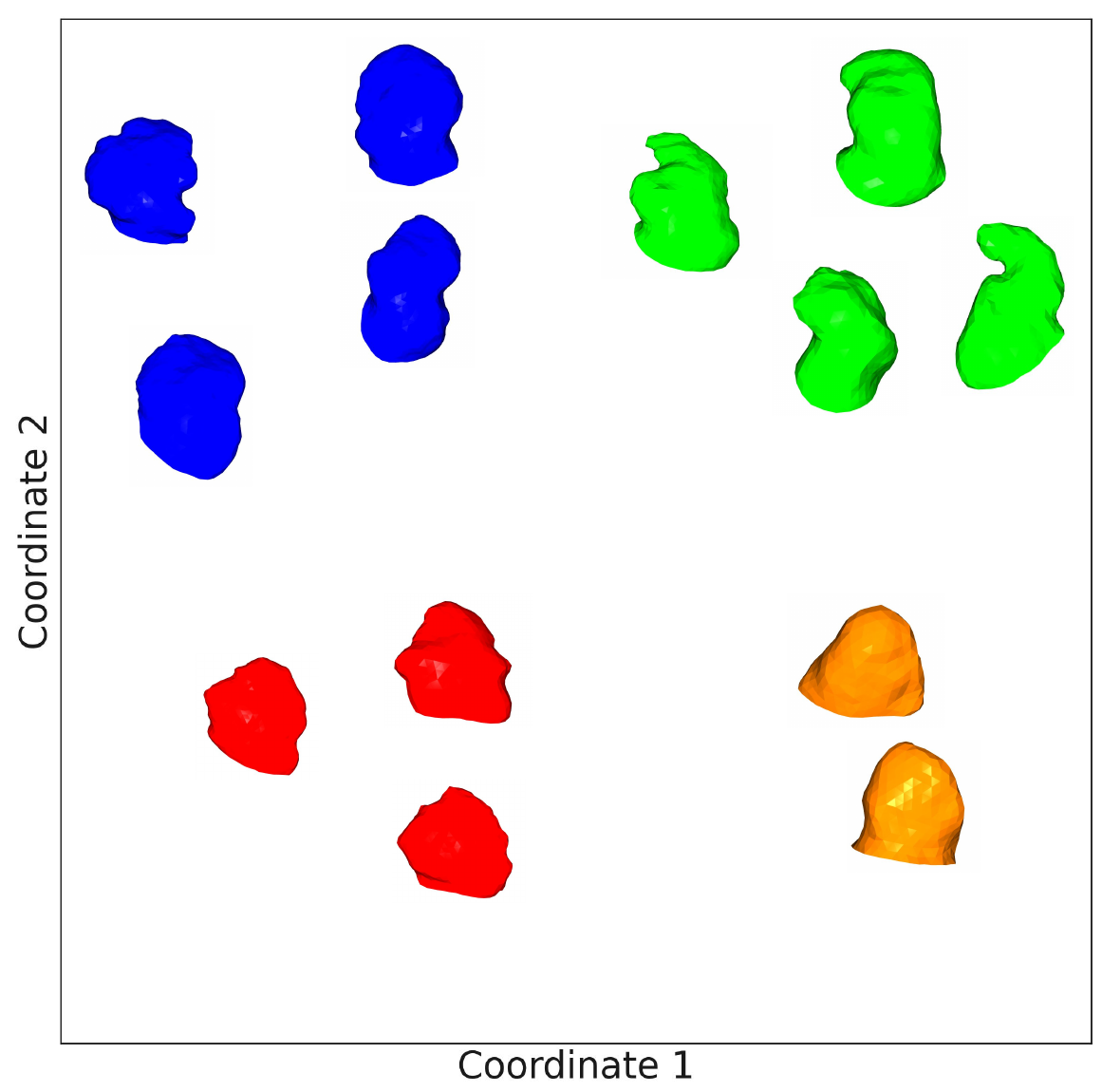}
        \caption{\textbf{LAA shape clusters.} The LAA meshes were projected into two-dimensional Euclidean space $\R^2$ with MDS, and the meshes are overlayed on the locations of their respective projections in $\R^2$. The LAA shapes are color-coded by their cluster labels, which were obtained from the $K$-means algorithm and agglomerative clustering with average linkage. To avoid overcrowding in the figure while providing a clear view of the geometry of the LAA meshes, we depict $13$ of the $20$ LAA meshes from our cohort, for which the clustering assignment was consistent (up to a permutation of the cluster labels) when using both of the aforementioned clustering algorithms.}
        \label{fig:laa_clusters}
\end{figure}

The aforementioned observation allows us to group the LAA meshes based on their shapes by leveraging their corresponding MDS projections and the Euclidean distances from Figure~\ref{fig:mdsDmatrix}. More specifically, we cluster our cohort of LAA meshes using standard clustering algorithms that operate directly in Euclidean space, including the $K$-means algorithm (see~\ref{app:kmeans}) and agglomerative (hierarchical) clustering with the average linkage method (see~\ref{app:agglo_clustering}). We display the clusters obtained using our approach in Figure~\ref{fig:laa_clusters}.

\begin{figure}[htbp]
    \centering
    \begin{minipage}[t]{0.485\textwidth}
        \centering
        \includegraphics[width=\textwidth]{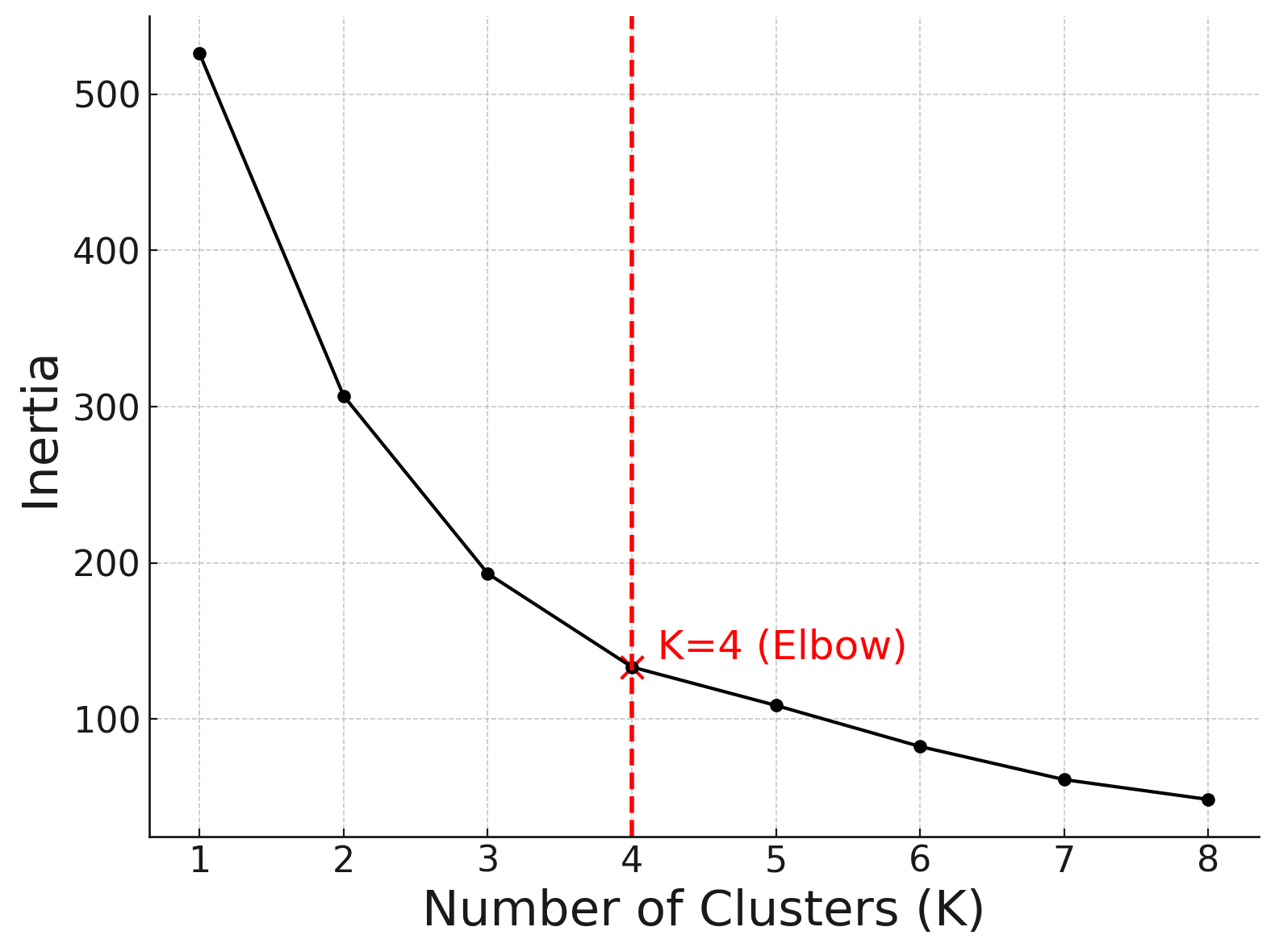}
        \subcaption{Elbow plot for $K$-means clustering.}
        \label{fig:elbow}
    \end{minipage}
    \hfill
    \begin{minipage}[t]{0.495\textwidth}
        \centering
        \vspace{-4.23cm} 
        \includegraphics[width=\textwidth]{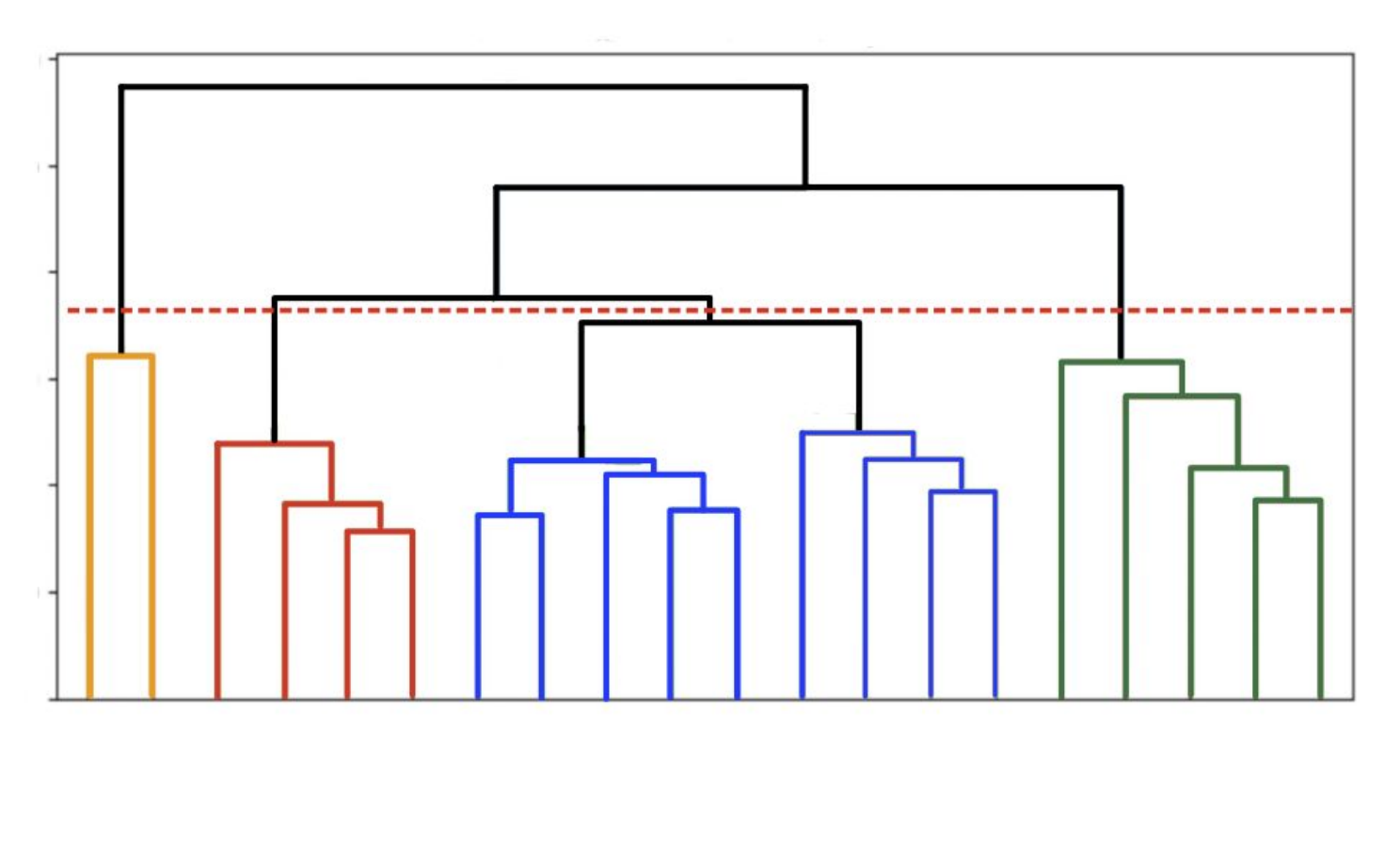}
        \subcaption{Dendrogram for agglomerative clustering results.}
        \label{fig:dendro}
    \end{minipage}
    \caption{\textbf{Left (a):} We plot the inertia~\eqref{eq:inertia} as a function of $K$, the number of clusters used for the $K$-means algorithm. Based on the inflection point of the inertia function, the optimal number of clusters is $K=4$. \textbf{Right (b):} We show the dendrogram from agglomerative clustering, which is color-coded to match the corresponding clusters in Figure~\ref{fig:laa_clusters}. The red dashed line indicates the threshold used to yield exactly four clusters.}
    \label{fig:elbow_dendro}
\end{figure}

The first thing to note is that our pipeline groups the set of LAA meshes into \emph{four} distinct clusters. The number of clusters was determined via the elbow method (see Figure~\ref{fig:elbow}). More specifically, we numerically computed the point of inflection (i.e., the so-called \emph{elbow}) in the graph of the inertia function~\eqref{eq:inertia}, which occurred at \mbox{\(K\!=\!4\)} (see~\ref{app:elbow-math} for further details). We therefore clustered our LAA dataset into \mbox{\(K\!=\!4\)} groups using both the $K$-means algorithm and agglomerative clustering (see Figure~\ref{fig:dendro}).

Strikingly, \mbox{$K=4$} corresponds to the number of qualitative categories of LAA morphologies (i.e., chicken wing, windsock, cactus, cauliflower) that was originally presented in the clinical work of Wunderlich et al.~\cite{wunderlich2015percutaneous}, and which is commonly used in clinical settings today. Furthermore, from a qualitative standpoint, the clusters obtained in Figure~\ref{fig:laa_clusters} demonstrate that our pipeline assigns LAAs with similar morphologies and geometric features into the same group, while separating LAAs with different shapes into distinct clusters. Notably, ``chicken wing'' appendages (the green meshes in Figure~\ref{fig:laa_clusters}) are grouped together, while ``non-chicken wing" LAAs are assigned to different clusters. The remarkable agreement between the LAA shape clusters determined by our pipeline and current qualitative LAA classification systems provides strong support for the relevance and potential applicability of our method in clinical contexts.

\subsection{Quantitative analysis of clustering results}
\label{ssec:quant_analysis_clusering}
To further analyze the LAA shape categorization produced by our pipeline, we provide a quantitative evaluation of our clustering results. To begin with, we report a series of cluster quality scores in Table~\ref{tab:clustering_results}, which were computed for the clusters produced via both the $K$-means algorithm and agglomerative clustering. In particular, we report:

\begin{itemize}
    \item[(i)] The \emph{silhouette score}~\cite{rousseeuw1987silhouettes}, which quantifies intra-cluster similarity compared to inter-cluster dissimilarity across all data points (see \ref{app:silhouette}). The silhouette score lies between $[-1,1]$, with higher values indicating better defined clusters characterized by strong intra-cluster cohesion and high inter-cluster separation.
    \item[(ii)] The \emph{Calinski-Harabasz Index (CHI)}~\cite{calinski1974dendrite}, which measures the (normalized) ratio of between-cluster separation to within-cluster dispersion across the dataset (see \ref{app:CH_index}). The CHI gives well-formed clusters higher values, and ill-formed clusters lower ones.
    \item[(iii)] The \emph{Davies-Bouldin Index (DBI)}~\cite{davies1979cluster}, which evaluates the average similarity between each cluster and its most similar neighboring cluster (see \ref{app:DBindex}). A lower DBI value indicates well-separated clusters with minimal overlap. A strong DBI score would be less than $1$, while a poor score would be greater than $1.5$.
\end{itemize}

\begin{table}[H]
\centering
\begin{tabular}{lcc}
\toprule
\textbf{Metric} & \textbf{$K$-means ($K=4$)} & \textbf{Agglomerative} \\
\midrule
Silhouette Score & $~0.67 \pm 0.024$ & $~0.64$ \\
Calinski-Harabasz Index & $26.72 \pm 0.360$ & $25.32$ \\
Davies-Bouldin Index & $~0.45 \pm 0.013$ & $~0.43$ \\
\bottomrule
\end{tabular}
\caption{Clustering quality metrics for the clusters produced by our pipeline using the $K$-means algorithm and agglomerative clustering (with average linkage).}
\label{tab:clustering_results}
\end{table}

As outlined above and in \ref{app:cluster_quality_scores}, the values of the metrics in Table~\ref{tab:clustering_results} indicate that the clusters produced by our pipeline are typified by strong levels of intra-cluster similarity, and high levels of inter-cluster separation. These metrics therefore provide quantitative validation for the observation that LAAs with similar shapes are grouped together in Figure~\ref{fig:laa_clusters}, while LAAs with distinct shapes are assigned to different groups.

\begin{remark}
On a technical note, the clustering assignment produced by the $K$-means algorithm exhibits a (slight) dependence on the random initialization of cluster centroids required for the procedure. To mitigate this issue, we ran $K$-means \(100\) times, and computed the cluster quality scores using the clusters obtained for each run. The mean and \(95\%\) confidence intervals of these scores across the $100$ runs are displayed in Table~\ref{tab:clustering_results}.
\end{remark}

Moreover, given that we used both the $K$-means and agglomerative clustering algorithms, it is important to assess the robustness and consistency of LAA clusters obtained from these different methods. To do so, we report the \emph{Adjusted Rand Index (ARI)}~\cite{halkidi2001clustering} and the \emph{Normalized Mutual Information (NMI)}~\cite{danon2005comparing}, which are clustering consistency metrics that measure the similarity between a pair of clustering assignments (see~\ref{app:cluster_consistency_scores}). The closer these scores are to $1$, the higher the degree of similarity between clustering assignments.

In particular, the ARI was \(0.908 \!\pm \!0.029\), and the NMI was \(0.897 \!\pm \!0.034\), indicating a high degree of overlap between the assignments obtained from the two clustering methods. Again, these $95\%$ confidence intervals are from repeatedly computing the $K$-means clusters. The high consistency from the ARI and NMI scores suggests that the categories resulting from our pipeline are not artifacts of a selected clustering method. Instead, the LAA groups obtained are reflective of intrinsic differences in the geometric features and shapes of LAAs across subjects, which are accurately captured by the elastic shape distances used in our pipeline.

\section{Discussion}
\label{sec:discussion}
We now provide a discussion our LAA shape categorization pipeline, and comment on its applicability in clinical settings.

\subsection{Advantages and clinical significance}
\label{ssec:advantages}
Unlike previous studies that necessitate precise point-to-point correspondences for comparing and categorizing LAA shapes, our pipeline is adapted for unparametrized surfaces, leveraging the relaxed matching approach in~\eqref{eq:relaxed_matching}.

By alleviating the need for point correspondences, our approach thus overcomes the computational hurdles and compatibility issues often encountered during the registration of complex geometric data such as LAA meshes \cite{slipsager2019statistical,juhl2024signed,ovsjanikov2012functional}.

Moreover, our framework's ability to operate on unparametrized surfaces enables us to work with LAA meshes with diverse vertex counts and discretization schemes, all with minimal pre-processing. Our pipeline can thus process data that has been acquired via different imaging modalities, and/or segmented via different techniques, and/or with data from different hospitals, which is of utmost practical value in clinical contexts.

\rev{In contrast, landmark-based approaches, such as the one used by Lee et al.~\cite{lee2024preserving}, require the selection and optimization of discrete point correspondences across samples, making them inherently sensitive to landmark placement. This reliance introduces variability depending on the chosen landmark distribution and optimization parameters, potentially affecting the stability and interpretability of shape comparisons.  In contrast, measure-based approaches such as varifold representations operate on continuous distributions of shape information, making them more robust to discretization differences and capable of capturing geometric variability without requiring explicit landmark alignment. By leveraging a relaxed matching framework, our approach avoids the limitations of correspondence-based methods and provides a more flexible and generalizable analysis of LAA morphology.}

\subsection{Limitations and future work}
\label{ssec:limitations}
Nevertheless, our work does possess certain limitations. Most prominently, our proof-of-concept study was conducted on a limited cohort of LAA meshes from 20 patients. As a result, a more in-depth statistical analysis of our results was not possible at this time. A follow-up study is underway, in which the authors are acquiring a larger LAA dataset with patient stroke history information, where the goal will be to refine the shape categorization system and identify clusters with higher stroke prevalence. This effort would allow clinicians to objectively compare new LAA geometries—sourced by various imaging acquisition modalities and healthcare centers— into one of the predefined groups, or to measure their distance from high-risk groups. This could lead to new metrics for stroke likelihood in AF patients and improve clinical decision-making regarding the administration of preventative medications (such as anticoagulants) with severe side effects \cite{steiner2006intracerebral}.

Moreover, from a computational perspective, the main bottleneck in our pipeline is the computation time required for pairwise distance calculations on large datasets. At the moment, it takes approximately $1-3$ minutes (on a GPU) to calculate the elastic shape distance between a pair of LAA meshes, mainly due to the complexity of the relaxed matching problem~\eqref{eq:relaxed_matching}. An avenue for addressing this limitation involves adapting the supervised learning approach of~\cite{hartman2021supervised}, or the self-supervised approach of~\cite{hartman2023varigrad}, for fast computations of the elastic distances via a forward pass through an appropriately trained neural network.

Additionally, parameter selection for the elastic distance computations can be intricate. Indeed, calculating the elastic distances entails choosing the coefficients of the $H^2$-metric~\eqref{eq:H2metric}, which we currently perform by grid search followed with sequential parameter refinement. As an avenue for improvement, one can envisage adapting the data-driven approach of~\cite{bauer2024elastic} to refine the parameter selection procedure. We note, however, that these parameter tuning tasks are typically one-time procedures that are essential for establishing robust shape categories. 

Additionally, we have not yet validated the elastic distances against other shape dissimilarity metrics, such as LDDMM~\cite{beg2005computing}, due to the limited sample size of LAA geometries available to us at this time. Future research will focus on developing faster methods for distance computations and expanding our analysis to a larger dataset annotated with stroke history. This will allow us to benchmark various deformation-based distance computation techniques to further validate the consistency of our clusters. 
\section{Conclusion} 
\label{sec:conclusion}
We have presented a comprehensive computational pipeline for LAA shape categorization, with applications towards improved stroke risk management in AF patients. Our approach, which combines elastic shape analysis with unsupervised learning for the categorization of LAA morphology in AF patients, yields robust LAA shape clusters which are reflective of the complex geometric variations of LAA shapes across subjects. The strong alignment between the LAA shape clusters identified by our pipeline and existing qualitative LAA classification systems highlights the relevance and potential clinical applicability of our method, emphasizing its significance in managing stroke risk for patients with atrial fibrillation. \\

\noindent
\textbf{Data Availability:} All de-identified data, 
\texttt{Python} code and specific dependencies / packages are available upon request for reproducibility.

\noindent
\textbf{Acknowledgments:} Z.A. was supported by grant n. 24PRE1196125 - American Heart Association (AHA) predoctoral fellowship and grant n. T32 HL007024 from the National Heart, Lung, and Blood Institute. M.Y. would like to acknowledge support from Heart Rhythm Society Fellowship. N.A.T. acknowledges National Institutes of Health (NIH) grants n. R01HL166759 and R01HL142496 and the Leducq Foundation. 

\noindent
\textbf{Author Contributions:} Z.A., M.Y., Y.S.: Conceptualization, methodology, software implementation, data preprocessing, simulation, formal analysis, writing (original draft). N.R.: Data preprocessing, formal analysis. 
E.K.: Data acquisition and processing, supervision, formal analysis, writing (review).
N.T.: Funding acquisition, conceptualization, supervision, project administration, formal analysis, writing (review).

\appendix

\section{Clustering algorithms}
\label{app:clustering_algos}
We provide brief summaries of the algorithms used as part of our LAA clustering pipeline.

\subsection{Multidimensional scaling}
\label{app:mds}
We employ multidimensional scaling (MDS)~\cite{torgerson1952multidimensional} to project the LAA meshes as points in Euclidean space while preserving relative information about their pairwise distances in the original shape space. More specifically, given a set of LAA shapes \(\{q_1,\dots,q_n\} \in \Shape\) with associated geodesic distance function \(d_{\Shape}\), MDS finds their corresponding projections \(\widehat{x}_1, \dots, \widehat{x}_n \in \R^m\) (for some \(m \in \N\)) by solving:
\begin{equation}
    \label{eq:mds_optimization}
    \widehat{x}_1, \ldots, \widehat{x}_n = \underset{y_1,\ldots,y_n \in \R^m}{\argmin} \Big( \sum_{i \neq j}  (d_{\Shape}(q_i, q_j) - \|y_i - y_j\|)^2 \Big)^{\frac{1}{2}}.
\end{equation}
In the results presented in Section~\ref{sec:results}, we chose \(m=2\).

\subsection{$K$-means clustering}
\label{app:kmeans}
We utilized the \(K\)-means clustering algorithm to cluster the projected LAA meshes. This algorithm aims to partition data points into \(K\) non-overlapping clusters by minimizing the sum of squared Euclidean distances between each data point and the centroid of its assigned cluster. Formally, given a set of points \(\{\widehat{x}_1, \dots, \widehat{x}_n\} \in \R^m\) (such as those obtained by projecting the LAA meshes into \(\R^2\) via MDS), the \(K\)-means algorithm finds clusters \(\widehat{C}_{1}, \dots, \widehat{C}_{K}\) by solving the following optimization problem:
\begin{equation}
    \label{eq:k_means_optimization}
    \widehat{C}_{1},\dots,\widehat{C}_{K} = \underset{C_{1}, \dots, C_{K}}{\argmin} ~\sum_{k=1}^{K} \sum_{\widehat{x} \in C_{k}} \|\widehat{x} - \mu_{k}\|^2,
\end{equation}
where \(\widehat{x}\) is a data point in cluster \(C_{k}\), and \(\mu_{k}\) is the centroid (mean) of cluster \(C_{k}\), defined as:
\begin{equation}
    \label{eq:centroid_kmeans}
    \mu_{k} = \frac{1}{|C_{k}|} \sum_{\widehat{x} \in C_{k}} \widehat{x},
\end{equation}
with \(|C_{k}|\) denoting the number of points in cluster \(C_{k}\), and \(\| \widehat{x} - \mu_{k} \|^2\) representing the squared Euclidean distance between \(\widehat{x}\) and \(\mu_{k}\). The algorithm iteratively assigns points to the cluster corresponding to the nearest centroid, updates the centroids, and repeats until convergence.

\subsubsection{Determining the optimal number of clusters}
\label{app:elbow-math}
A key component of the \(K\)-means algorithm entails choosing the optimal number of clusters \(K\) for partitioning the dataset. To do so, we leverage the so-called \emph{inertia function} \(I(K)\), which is simply the loss function being minimized in~\eqref{eq:k_means_optimization}, when viewed as a function of the number of clusters:
\begin{equation}
    \label{eq:inertia}
    I(K) = \sum_{k=1}^{K} \sum_{\widehat{x} \in C_{k}} \| \widehat{x} - \mu_{k} \|^2.
\end{equation}
The inertia \(I(K)\) decreases as \(K\) increases, due to clusters containing fewer members, and more points being closer to their centroids. However, after a certain point, increasing \(K\) only results in small reductions in the inertia, indicating that further partitioning does not significantly improve the clustering quality, as seen in Figure~\ref{fig:elbow}.

We determined the point of inflection in Figure~\ref{fig:elbow} (the so-called \emph{elbow}) by finding the point at which the second derivative of the inertia function (with respect to the number of clusters \(K\)) was maximized in absolute value. For our dataset of LAA meshes, we found that the elbow occurred at \(K=4\), and thus clustered our LAA dataset into \(K=4\) groups.

\subsection{Agglomerative clustering}
\label{app:agglo_clustering}
We also applied agglomerative clustering to categorize our LAA meshes. Agglomerative clustering is a hierarchical clustering technique that builds a nested hierarchy of clusters by starting with each data point as its own cluster and repeatedly merging the most similar clusters. Given a set of points \(\{\widehat{x}_1, \dots, \widehat{x}_n\} \in \R^m\), with pairwise distances \(D_{ij} = \|\widehat{x}_i - \widehat{x}_j\|\) for $i,j=1,\dots,n$, agglomerative clustering proceeds by minimizing a linkage criterion at each iteration. We use the \textit{average linkage} method below:
\begin{equation}
    \label{eq:average_linkage_agglo_clustering}
    d(C_{k}, C_{\ell}) = \frac{1}{|C_{k}| |C_{\ell}|} \sum_{\widehat{x} \in C_{k}} \sum_{\widehat{y} \in C_{\ell}} \|\widehat{x} - \widehat{y}\|,
\end{equation}
where \(C_{k}\) and \(C_{\ell}\) denote two different clusters being merged. At each step, the pair of clusters \(C_{k}\) and \(C_{\ell}\) that minimizes the selected linkage criterion is merged, and the process repeats until the desired number of clusters is obtained.  

\section{Cluster quality scores}
\label{app:cluster_quality_scores}
After applying the aforementioned clustering algorithms to cluster our projected LAA meshes, we quantitatively evaluated the quality of the resulting clusters using a variety of metrics, which we detail below.

\subsection{Silhouette score}
\label{app:silhouette}
The first metric we employed is the \emph{silhouette score}~\cite{rousseeuw1987silhouettes}, which measures the similarity of each data point to its own cluster (intra-cluster similarity) compared to the nearest neighboring cluster (inter-cluster dissimilarity). Indeed, given a set of points \(\{\widehat{x}_1, \dots, \widehat{x}_n\} \in \R^m\) (such as the set of LAA meshes projected into \(\R^2\) via MDS) which have been assigned to $K$ clusters $C_1,\dots, C_K$, the silhouette score of data point \(\widehat{x}_i\), denoted by  \(s(\widehat{x}_i)\) for each $i=1,\dots,n$, is defined as:
\begin{equation}
    \label{eq:silhouette_score_datapoint}
    s(\widehat{x}_i) = \frac{b(\widehat{x}_i) - a(\widehat{x}_i)}{\max\{a(\widehat{x}_i), b(\widehat{x}_i)\}},
\end{equation}
where:
\begin{itemize}
    \item \(a(\widehat{x}_i) = \frac{1}{|C| - 1} \sum_{\widehat{x}_j \in C, ~j \neq i} \|\widehat{x}_i - \widehat{x}_j\|\) is the mean intra-cluster distance for data point \(\widehat{x}_i\), and \(C\) is the cluster to which \(\widehat{x}_i\) belongs.
    \item \(b(\widehat{x}_i) = \underset{C_k \neq C}{\min} ~\frac{1}{|C_k|} \sum_{\widehat{x}_j \in C_k} \|\widehat{x}_i - \widehat{x}_j\|\) is the mean distance between \(\widehat{x}_i\) and all points in the nearest neighboring cluster \(C_k\), where \(k\) is chosen to minimize the mean distance.
\end{itemize}
The overall silhouette score \(s\) for the dataset is the average silhouette score across all data points:
\begin{equation}
    \label{eq:silhouette_score_dataset}
    s = \frac{1}{n} \sum_{i=1}^n s(\widehat{x}_i).
\end{equation}
The silhouette score $s$ lies in the range \([-1, 1]\), with higher values indicating better-defined clusters. A higher score implies strong intra-cluster cohesion and inter-cluster separation.

\subsection{Calinski-Harabasz Index (CHI)}
\label{app:CH_index}
The next metric we used is the \emph{Calinski-Harabasz index (CHI)}~\cite{calinski1974dendrite}. Given a set of data points \(\{\widehat{x}_1, \dots, \widehat{x}_n\} \in \R^m\) that have been assigned to $K$ clusters $C_1,\dots, C_K$, the $\operatorname{CHI}$ is defined as the ratio of the between-cluster separation $B$ to the within-cluster dispersion $W$, normalized by their number of degrees of freedom:
\begin{equation}
    \label{eq:ch_index}
    \operatorname{CHI} = \frac{B / (K - 1)}{W / (n - K)},
\end{equation}
where:
\begin{itemize}
    \item \(B = \sum_{k=1}^K |C_k| \|\mu_k - \mu\|^2\), with \(\mu_k\) being the centroid of cluster \(C_k\) and \(\mu\) being the global centroid of all data points.
    \item \(W = \sum_{k=1}^K \sum_{\widehat{x}_i \in C_k} \|\widehat{x}_i - \mu_k\|^2\).
\end{itemize}
A higher $\operatorname{CHI}$ index value indicates better-defined clusters, as it reflects a higher ratio of between-cluster dispersion to within-cluster dispersion.

\subsection{Davies-Bouldin Index (DBI)}
\label{app:DBindex}
Lastly, we used the \emph{Davies-Bouldin index (DBI)}~\cite{davies1979cluster}, which evaluates the average similarity between each cluster and its most similar neighboring cluster. It is defined as:
\begin{equation}
    \label{eq:db_index}
    \operatorname{DBI} = \frac{1}{K} \sum_{k=1}^{K} \max_{\ell \neq k} \frac{d_k + d_\ell}{d_{k\ell}},
\end{equation}
where:
\begin{itemize}
    \item \(d_k = \frac{1}{|C_k|} \sum_{\widehat{x} \in C_k} \|\widehat{x} - \mu_k\|\) is the average intra-cluster distance for cluster \(C_k\), with \(\mu_k\) as the centroid of cluster \(C_k\).
    \item \(d_{k\ell} = \|\mu_k - \mu_\ell\|\) is the Euclidean distance between the centroids of clusters \(C_k\) and \(C_\ell\).
\end{itemize}
The $\operatorname{DBI}$ lies in the range \([0, \infty)\), with lower values indicating better separation between clusters. A strong $\operatorname{DBI}$ score would be less than 1, while a poor score would be greater than 1.5. A lower $\operatorname{DBI}$ value indicates well-separated clusters with minimal overlap.

\section{Cluster consistency scores}
\label{app:cluster_consistency_scores}
Given that we used various clustering algorithms to categorize the projected LAA data points, an important consideration involves assessing the robustness and consistency of clusters obtained from the different clustering methods. In this section, we outline the metrics used to quantitatively compare clustering assignments across different algorithms.

\subsection{Adjusted Rand Index (ARI)}
\label{app:ARI}
First, we used the \emph{Adjusted Rand Index (ARI)}~\cite{halkidi2001clustering}, which measures the similarity between a pair of clustering assignments, adjusted for chance. Consider a set of data points \(\{\widehat{x}_1, \dots, \widehat{x}_n\} \in \R^m\), and a pair of clustering assignments \( A = \{A_k\}_{k=1}^K \) and \( B = \{B_{\ell}\}_{\ell=1}^L \) for these data points, obtained for instance from two clustering algorithms. The $\operatorname{ARI}$ is given by:
\begin{equation}
    \label{eq:ari}
    \text{ARI} = \frac{\sum_{k,\ell} \binom{n_{k\ell}}{2} - \left[ \sum_{k} \binom{a_k}{2} \sum_{\ell} \binom{b_\ell}{2} \right] / \binom{n}{2}}{\frac{1}{2} \left[ \sum_{k} \binom{a_k}{2} + \sum_{\ell} \binom{b_\ell}{2} \right] - \left[ \sum_{k} \binom{a_k}{2} \sum_{\ell} \binom{b_\ell}{2} \right] / \binom{n}{2}},
\end{equation}
where:
\begin{itemize}
    \item \(n_{k\ell}\) is the number of points assigned to both cluster \(A_k \in A\) and \(B_\ell \in B\),
    \item \(a_k\) is the total number of points in cluster \(A_k\),
    \item \(b_\ell\) is the total number of points in cluster \(B_\ell\).
\end{itemize}
The $\operatorname{ARI}$ ranges from \([-1, 1]\), where higher values indicate better agreement between the two clustering solutions.

\subsection{Normalized Mutual Information (NMI)}
\label{app:NMI}
We also utilized the \emph{Normalized Mutual Information (NMI)}~\cite{danon2005comparing} to quantify the mutual dependence between two clustering solutions. Let \( U = \{U_k\}_{k=1}^K \) and \( V = \{V_\ell\}_{\ell=1}^L \) represent the two clusterings being compared. The $\operatorname{NMI}$ is given by:
\begin{equation}
    \label{eq:nmi}
    \text{NMI}(U, V) = \frac{2 I(U, V)}{H(U) + H(V)},
\end{equation}
where:
\begin{itemize}
    \item \(I(U, V) = \sum_{k, \ell} P(k, \ell) \log \frac{P(k, \ell)}{P(k) P(\ell)}\) is the mutual information between clusters \(U\) and \(V\),
    \item \(P(k, \ell)\) is the empirical joint probability of a point being in cluster \(U_k\) in clustering \(U\) and cluster \(V_\ell\) in clustering \(V\),
    \item \(P(k) = \frac{|U_k|}{n}\) and \(P(\ell) = \frac{|V_\ell|}{n}\) are the empirical probabilities of a point being in clusters \(U_k\) and \(V_\ell\), respectively,
    \item \(H(U) = -\sum_{k} P(k) \log P(k)\) and \(H(V) = -\sum_{\ell} P(\ell) \log P(\ell)\) are the entropies of the clusterings \(U\) and \(V\).
\end{itemize}

The $\operatorname{NMI}$ score lies in the range \([0, 1]\), with higher values indicating stronger similarity between clustering solutions.

These consistency metrics help validate whether different clustering methods yield comparable results, ensuring that the identified clusters are not artifacts of specific algorithms.

\bibliographystyle{elsarticle-num} 
\bibliography{mybibliography.bib}

\begin{thebibliography}{10}
\expandafter\ifx\csname url\endcsname\relax
  \def\url#1{\texttt{#1}}\fi
\expandafter\ifx\csname urlprefix\endcsname\relax\def\urlprefix{URL }\fi
\expandafter\ifx\csname href\endcsname\relax
  \def\href#1#2{#2} \def\path#1{#1}\fi

\bibitem{january20192019}
C.~T. January, L.~S. Wann, H.~Calkins, L.~Y. Chen, J.~E. Cigarroa, J.~C. Cleveland~Jr, P.~T. Ellinor, M.~D. Ezekowitz, M.~E. Field, K.~L. Furie, et~al., 2019 aha/acc/hrs focused update of the 2014 aha/acc/hrs guideline for the management of patients with atrial fibrillation: a report of the american college of cardiology/american heart association task force on clinical practice guidelines and the heart rhythm society in collaboration with the society of thoracic surgeons, Circulation 140~(2) (2019) e125--e151.

\bibitem{pistoia2016epidemiology}
F.~Pistoia, S.~Sacco, C.~Tiseo, D.~Degan, R.~Ornello, A.~Carolei, The epidemiology of atrial fibrillation and stroke, Cardiology clinics 34~(2) (2016) 255--268.

\bibitem{yang2023observational}
Q.~Yang, S.~Liu, J.~Wang, Y.~Guo, An observational study: Clinical manifestations and prognosis of left atrial thrombosis in atrial fibrillation, SN Comprehensive Clinical Medicine 5~(1) (2023) 159.

\bibitem{wolf1991atrial}
P.~A. Wolf, R.~D. Abbott, W.~B. Kannel, Atrial fibrillation as an independent risk factor for stroke: the framingham study., stroke 22~(8) (1991) 983--988.

\bibitem{piccini2016preventing}
J.~P. Piccini, G.~C. Fonarow, Preventing stroke in patients with atrial fibrillation—a steep climb away from achieving peak performance, JAMA cardiology 1~(1) (2016) 63--64.

\bibitem{tian2020morphological}
X.~Tian, X.-J. Zhang, Y.-F. Yuan, C.-Y. Li, L.-X. Zhou, B.-L. Gao, Morphological and functional parameters of left atrial appendage play a greater role in atrial fibrillation relapse after radiofrequency ablation, Scientific Reports 10~(1) (2020) 8072.

\bibitem{bieging2021statistical}
E.~T. Bieging, A.~Morris, L.~Chang, L.~Dagher, N.~F. Marrouche, J.~Cates, Statistical shape analysis of the left atrial appendage predicts stroke in atrial fibrillation, The international journal of cardiovascular imaging 37~(8) (2021) 2521--2527.

\bibitem{burrell2013usefulness}
L.~D. Burrell, B.~D. Horne, J.~L. Anderson, J.~B. Muhlestein, B.~K. Whisenant, Usefulness of left atrial appendage volume as a predictor of embolic stroke in patients with atrial fibrillation, The American journal of cardiology 112~(8) (2013) 1148--1152.

\bibitem{di2012does}
L.~Di~Biase, P.~Santangeli, M.~Anselmino, P.~Mohanty, I.~Salvetti, S.~Gili, R.~Horton, J.~E. Sanchez, R.~Bai, S.~Mohanty, et~al., Does the left atrial appendage morphology correlate with the risk of stroke in patients with atrial fibrillation? results from a multicenter study, Journal of the American College of Cardiology 60~(6) (2012) 531--538.

\bibitem{harb2018p5142}
S.~Harb, A.~Hussein, W.~Saliba, Y.~Wu, B.~Xu, L.~Cho, O.~Wazni, W.~Jaber, P5142 effect of anticoagulation on mortality by chadsvasc score in patients with atrial fibrillation: comparison to patients without atrial fibrillation, European Heart Journal 39~(suppl\_1) (2018) ehy566--P5142.

\bibitem{steiner2006intracerebral}
T.~Steiner, J.~Rosand, M.~Diringer, Intracerebral hemorrhage associated with oral anticoagulant therapy: current practices and unresolved questions, Stroke 37~(1) (2006) 256--262.

\bibitem{chen2019cha2ds2}
L.~Y. Chen, F.~L. Norby, A.~M. Chamberlain, R.~F. MacLehose, L.~G. Bengtson, P.~L. Lutsey, A.~Alonso, Cha2ds2-vasc score and stroke prediction in atrial fibrillation in whites, blacks, and hispanics, Stroke 50~(1) (2019) 28--33.

\bibitem{zhang2021interpretation}
J.~Zhang, R.~Lenarczyk, F.~Marin, K.~Malaczynska-Rajpold, J.~Kosiuk, W.~Doehner, I.~C. Van~Gelder, G.~Lee, J.~M. Hendriks, G.~Y. Lip, et~al., The interpretation of cha2ds2-vasc score components in clinical practice: a joint survey by the european heart rhythm association (ehra) scientific initiatives committee, the ehra young electrophysiologists, the association of cardiovascular nursing and allied professionals, and the european society of cardiology council on stroke, EP Europace 23~(2) (2021) 314--322.

\bibitem{dudzinska2022association}
K.~Dudzi{\~n}ska-Szczerba, P.~Ku{\l}akowski, I.~Micha{\l}owska, J.~Baran, Association between left atrial appendage morphology and function and the risk of ischaemic stroke in patients with atrial fibrillation, Arrhythmia \& Electrophysiology Review 11 (2022).

\bibitem{sun2023finding}
Y.~Sun, Y.~Ling, Z.~Chen, Z.~Wang, T.~Li, Q.~Tong, Y.~Qian, Finding low cha2ds2-vasc scores unreliable? why not give morphological and hemodynamic methods a try?, Frontiers in Cardiovascular Medicine 9 (2023) 1032736.

\bibitem{lodzinski2020trends}
P.~Lodzi{\'n}ski, M.~Gawa{\l}ko, M.~Budnik, A.~Tymi{\'n}ska, K.~Oziera{\'n}ski, M.~Grabowski, A.~Janion-Sadowska, G.~Opolski, R.~Lenarczyk, Z.~Kalarus, et~al., Trends in antithrombotic management of patients with atrial fibrillation: A report from the polish part of the eurobservational research programme-atrial fibrillation general long-term registry, Polskie Archiwum Medycyny Wewnetrznej 130~(3) (2020) 196--205.

\bibitem{fang2022stroke}
R.~Fang, Y.~Li, J.~Wang, Z.~Wang, J.~Allen, C.~K. Ching, L.~Zhong, Z.~Li, Stroke risk evaluation for patients with atrial fibrillation: Insights from left atrial appendage, Frontiers in Cardiovascular Medicine 9 (2022) 968630.

\bibitem{smit2021anatomical}
J.~M. Smit, J.~Simon, M.~El~Mahdiui, L.~Szaraz, P.~J. van Rosendael, M.~Kolassv{\'a}ry, B.~Szilveszter, V.~Delgado, B.~Merkely, P.~Maurovich-Horvat, et~al., Anatomical characteristics of the left atrium and left atrial appendage in relation to the risk of stroke in patients with versus without atrial fibrillation, Circulation: Arrhythmia and Electrophysiology 14~(8) (2021) e009777.

\bibitem{yaghi2020left}
S.~Yaghi, A.~D. Chang, R.~Akiki, S.~Collins, T.~Novack, M.~Hemendinger, A.~Schomer, B.~Mac~Grory, S.~Cutting, T.~Burton, et~al., The left atrial appendage morphology is associated with embolic stroke subtypes using a simple classification system: a proof of concept study, Journal of cardiovascular computed tomography 14~(1) (2020) 27--33.

\bibitem{zingaro2024comprehensive}
A.~Zingaro, Z.~Ahmad, E.~Kholmovski, K.~Sakata, L.~Dede’, A.~K. Morris, A.~Quarteroni, N.~A. Trayanova, A comprehensive stroke risk assessment by combining atrial computational fluid dynamics simulations and functional patient data, Scientific reports 14~(1) (2024) 9515.

\bibitem{zingaro2023po}
A.~Zingaro, Z.~Ahmad, C.~Y.~A. Pinto, K.~Sakata, E.~G. Kholmovski, A.~Quarteroni, N.~A. Trayanova, et~al., Po-01-210 stroke risk is identified by slow blood flow and stagnant blood particles in the left atrium, Heart Rhythm 20~(5) (2023) S161--S162.

\bibitem{slodowska2021morphology}
K.~S{\l}odowska, E.~Szczepanek, D.~Dudkiewicz, J.~Ho{\l}da, F.~Bolecha{\l}a, M.~Strona, M.~Lis, J.~Batko, M.~Koziej, M.~K. Ho{\l}da, Morphology of the left atrial appendage: introduction of a new simplified shape-based classification system, Heart, Lung and Circulation 30~(7) (2021) 1014--1022.

\bibitem{slipsager2019statistical}
J.~M. Slipsager, K.~A. Juhl, P.~E. Sigvardsen, K.~F. Kofoed, O.~De~Backer, A.~L. Olivares, O.~Camara, R.~R. Paulsen, Statistical shape clustering of left atrial appendages, in: Statistical Atlases and Computational Models of the Heart. Atrial Segmentation and LV Quantification Challenges: 9th International Workshop, STACOM 2018, Held in Conjunction with MICCAI 2018, Granada, Spain, September 16, 2018, Revised Selected Papers 9, Springer, 2019, pp. 32--39.

\bibitem{juhl2024signed}
K.~A. Juhl, J.~Slipsager, O.~de~Backer, K.~Kofoed, O.~Camara, R.~Paulsen, Signed distance field based segmentation and statistical shape modelling of the left atrial appendage, arXiv preprint arXiv:2402.07708 (2024).

\bibitem{lee2024preserving}
M.~T. Lee, V.~Martorana, R.~I. Md, R.~Sivera, A.~C. Cook, L.~Menezes, G.~Burriesci, R.~Torii, G.~M. Bosi, On preserving anatomical detail in statistical shape analysis for clustering: focus on left atrial appendage morphology, Frontiers in Network Physiology 4 (2024) 1467180.

\bibitem{cates2017shapeworks}
J.~Cates, S.~Elhabian, R.~Whitaker, Shapeworks: particle-based shape correspondence and visualization software, in: Statistical Shape and Deformation Analysis, Elsevier, 2017, pp. 257--298.

\bibitem{ovsjanikov2012functional}
M.~Ovsjanikov, M.~Ben-Chen, J.~Solomon, A.~Butscher, L.~Guibas, Functional maps: a flexible representation of maps between shapes, ACM Transactions on Graphics (ToG) 31~(4) (2012) 1--11.

\bibitem{lefebvre2022lassnet}
A.~L. Lefebvre, C.~A. Yamamoto, J.~K. Shade, R.~P. Bradley, R.~A. Yu, R.~L. Ali, D.~M. Popescu, A.~Prakosa, E.~G. Kholmovski, N.~A. Trayanova, Lassnet: A four steps deep neural network for left atrial segmentation and scar quantification, in: Challenge on Left Atrial and Scar Quantification and Segmentation, Springer, 2022, pp. 1--15.

\bibitem{schluchter2019vascular}
A.~Schluchter, C.~Jan, K.~Lowe, D.~M. Vigneault, F.~Contijoch, E.~R. McVeigh, Vascular landmark-based method for highly reproducible measurement of left atrial appendage volume in computed tomography, Circulation: Cardiovascular Imaging 12~(12) (2019) e009075.

\bibitem{hartman2023elastic}
E.~Hartman, Y.~Sukurdeep, E.~Klassen, N.~Charon, M.~Bauer, Elastic shape analysis of surfaces with second-order sobolev metrics: a comprehensive numerical framework, International Journal of Computer Vision 131~(5) (2023) 1183--1209.

\bibitem{bauer2012sobolev}
M.~Bauer, P.~Harms, P.~W. Michor, Sobolev metrics on shape space of surfaces, Journal of Geometric Mechanics 3~(4) (2012) 389--438.

\bibitem{jermyn2017elastic}
I.~H. Jermyn, S.~Kurtek, H.~Laga, A.~Srivastava, G.~Medioni, S.~Dickinson, Elastic shape analysis of three-dimensional objects, Springer, 2017.

\bibitem{beg2005computing}
M.~F. Beg, M.~I. Miller, A.~Trouv{\'e}, L.~Younes, Computing large deformation metric mappings via geodesic flows of diffeomorphisms, International journal of computer vision 61 (2005) 139--157.

\bibitem{su2020shape}
Z.~Su, M.~Bauer, S.~C. Preston, H.~Laga, E.~Klassen, Shape analysis of surfaces using general elastic metrics, Journal of Mathematical Imaging and Vision 62 (2020) 1087--1106.

\bibitem{bauer2021numerical}
M.~Bauer, N.~Charon, P.~Harms, H.-W. Hsieh, A numerical framework for elastic surface matching, comparison, and interpolation, International Journal of Computer Vision 129~(8) (2021) 2425--2444.

\bibitem{bauer2010sobolev}
M.~Bauer, P.~Harms, P.~W. Michor, Sobolev metrics on shape space of surfaces, arXiv preprint arXiv:1009.3616 (2010).

\bibitem{hartman2023bare}
E.~Hartman, E.~Pierson, M.~Bauer, N.~Charon, M.~Daoudi, Bare-esa: A riemannian framework for unregistered human body shapes, in: Proceedings of the IEEE/CVF International Conference on Computer Vision, 2023, pp. 14181--14191.

\bibitem{hartman2023basis}
E.~Hartman, E.~Pierson, M.~Bauer, M.~Daoudi, N.~Charon, Basis restricted elastic shape analysis on the space of unregistered surfaces, arXiv preprint arXiv:2311.04382 (2023).

\bibitem{kaltenmark2017general}
I.~Kaltenmark, B.~Charlier, N.~Charon, A general framework for curve and surface comparison and registration with oriented varifolds, in: Proceedings of the IEEE Conference on Computer Vision and Pattern Recognition, 2017, pp. 3346--3355.

\bibitem{charon2013varifold}
N.~Charon, A.~Trouv{\'e}, The varifold representation of nonoriented shapes for diffeomorphic registration, SIAM journal on Imaging Sciences 6~(4) (2013) 2547--2580.

\bibitem{tenenbaum2000global}
J.~B. Tenenbaum, V.~d. Silva, J.~C. Langford, A global geometric framework for nonlinear dimensionality reduction, science 290~(5500) (2000) 2319--2323.

\bibitem{wunderlich2015percutaneous}
N.~C. Wunderlich, R.~Beigel, M.~J. Swaans, S.~Y. Ho, R.~J. Siegel, Percutaneous interventions for left atrial appendage exclusion: options, assessment, and imaging using 2d and 3d echocardiography, JACC: Cardiovascular Imaging 8~(4) (2015) 472--488.

\bibitem{rousseeuw1987silhouettes}
P.~J. Rousseeuw, Silhouettes: a graphical aid to the interpretation and validation of cluster analysis, Journal of computational and applied mathematics 20 (1987) 53--65.

\bibitem{calinski1974dendrite}
T.~Cali{\'n}ski, J.~Harabasz, A dendrite method for cluster analysis, Communications in Statistics-theory and Methods 3~(1) (1974) 1--27.

\bibitem{davies1979cluster}
D.~L. Davies, D.~W. Bouldin, A cluster separation measure, IEEE transactions on pattern analysis and machine intelligence~(2) (1979) 224--227.

\bibitem{halkidi2001clustering}
M.~Halkidi, Y.~Batistakis, M.~Vazirgiannis, On clustering validation techniques, Journal of intelligent information systems 17 (2001) 107--145.

\bibitem{danon2005comparing}
L.~Danon, A.~Diaz-Guilera, J.~Duch, A.~Arenas, Comparing community structure identification, Journal of statistical mechanics: Theory and experiment 2005~(09) (2005) P09008.

\bibitem{hartman2021supervised}
E.~Hartman, Y.~Sukurdeep, N.~Charon, E.~Klassen, M.~Bauer, Supervised deep learning of elastic srv distances on the shape space of curves, in: Proceedings of the IEEE/CVF conference on computer vision and pattern recognition, 2021, pp. 4425--4433.

\bibitem{hartman2023varigrad}
E.~Hartman, E.~Pierson, Varigrad: A novel feature vector architecture for geometric deep learning on unregistered data, arXiv preprint arXiv:2307.03553 (2023).

\bibitem{bauer2024elastic}
M.~Bauer, N.~Charon, E.~Klassen, S.~Kurtek, T.~Needham, T.~Pierron, Elastic metrics on spaces of euclidean curves: Theory and algorithms, Journal of Nonlinear Science 34~(3) (2024) 1--37.

\bibitem{torgerson1952multidimensional}
W.~S. Torgerson, Multidimensional scaling: I. theory and method, Psychometrika 17~(4) (1952) 401--419.

\end{thebibliography}

\end{document}